# Polyvinylpyrrolidone Planarized Liquid Crystalline 1T-WS$_2$/rGO Hybrid Nanocomposites-based Humidity Sensing Platform

A. Vasil'ev[1], M. Zhezhu[1], S. Gyozalyan[1], L. Avanesyan[1], Y. Grigoryan[1], A.A. Kuzanyan[2], A. A. Hovhannisyan[3], Mohamed Aly Saad Aly[4,5], D. A. Ghazaryan[6], H. Gharagulyan[1, 7] *

[1] *Liquid Crystalline Nanosystems Research Group, A.B. Nalbandyan Institute of Chemical Physics NAS RA, Yerevan 0014, Armenia*

[2] *Institute for Physical Research National Academy of Sciences of Armenia, Ashtarak-2, 0204, Republic of Armenia*

[3] *The Scientific Technological Centre of Organic and Pharmaceutical Chemistry NAS RA, Molecule Structure Research Centre, 26 Azatutyan Av., Yerevan, 0014, Armenia*

[4] *School of Electrical and Computer Engineering, Georgia Institute of Technology, Atlanta, Georgia 30332, United States*

[5] *Department of Electrical and Computer Engineering at Georgia Tech Shenzhen Institute (GTSI), Shenzhen, Guangdong 518055, China*

[6] *Laboratory of Advanced Functional Materials, Yerevan State University, Yerevan 0025, Armenia*

[7] *Institute of Physics, Yerevan State University, Yerevan 0025, Armenia*

*Author to whom correspondence should be addressed: herminegharagulyan@ysu.am

## Abstract

Two-dimensional hybrid nanocomposites combining graphene-like materials and transition metal dichalcogenides (TMDCs) are created using various synthesis methods and are vital for environmental sensing due to the synergistic effects of their components. These hybrid materials offer enhanced sensitivity, selectivity, surface-to-volume ratio, tunable electronic properties, strong interaction with analytes, and fast detection capabilities, which address the limitations of individual materials. This is due to their outstanding surface-to-volume ratio, tunable electronic properties, strong interaction with analytes, as well as their high sensitivity, selectivity, and fast detection capabilities, which are crucial for environmental sensing. Herein, a hydrothermally synthesized and polyvinylpyrrolidone (PVP)-stabilized 1T phase tungsten disulfide/reduced graphene oxide, 1T-WS$_2$/PVP/rGO, hybrid nanocomposite is reported. The intrinsic liquid crystalline behavior of the synthesized composite enables the formation of highly uniform films for effective humidity sensing. The optostructural characterization and key performance parameters, including response and recovery times, of the 1T-WS$_2$/PVP/rGO hybrid nanocomposite-based structure are comprehensively analyzed and studied. This is the first to report on the synthesis of a 1T-WS$_2$/PVP/rGO hybrid nanocomposite, its liquid crystalline phase formation and application in humidity sensing. The proposed sensing platform introduces a novel approach to humidity sensing using 1T-WS$_2$/PVP/rGO liquid crystalline films, which combine the metallic advantages of 1T-WS$_2$, the stabilizing role of PVP and the conductive framework of rGO into aligned LC structures for enhanced sensitivity, rapid response and environmental robustness.

**Keywords**: 2D materials, WS$_2$, rGO, PVP, hybrid nanocomposites, liquid crystal, relative humidity, sensing.

## 1. Introduction

Transition metal dichalcogenides (TMDCs) are advanced two-dimensional (2D) materials renowned for their tunable electronic properties, strong light-matter interactions and layer-dependent bandgaps, which is making them ideal for next-generation nanoelectronics and quantum technologies [1-3]. Among them, tungsten disulfide (WS$_2$) stands out as a particularly promising candidate for advancing such technological applications [4, 5]. It can crystallize in various phases, such as 1T, 1T′, 2H and 3R while each exhibiting unique structural and electronic properties [6]. However, the most common polymorphs are 2H phase with semiconducting behavior and 1T phase with metallic characteristics [7]. 2H phase dominates WS$_2$ studies, reflecting its stability and broad applicability while 1T phase research is more growing area focusing on metallic and catalytic properties [8]. Alternatively, graphene derivatives offer distinct advantages, such as easy-to-scale synthesis, extremely high specific surface area, and mechanical robustness [9]. While GO is indeed highly hydrophilic due to its abundant oxygen-containing groups, which promote water adsorption and proton conduction in humidity sensors, its insulating nature limits its utility in resistive sensors requiring low baseline resistance and fast electro-optic response. In contrast, rGO restores partial conductivity through sp² network reconstruction while retaining sufficient residual oxygen groups evident by XPS analysis for moderate hydrophilicity. This makes rGO more suitable for application, where electronic conduction through a percolation network is essential for room-temperature sensitivity, alongside protonic contributions at high relative humidity (RH) [10]. In this context, combining TMDCs with reduced graphene oxide (rGO) into hybrid nanocomposites presents a strategically advantageous approach for tailoring materials to specific applications [11-13]. Notably, metallic 1T-WS$_2$ and rGO, with their respective conductive and surface area advantages work synergistically to deliver enhanced functionality beyond that of the individual materials.

The ability of these materials in their liquid-crystalline (LC) phases to form films with enhanced uniformity is crucial for applications requiring high performance and precise molecular arrangement [14-16]. These LC-aligned composites are particularly beneficial for optoelectronics, photonics and sensing devices requiring uniform morphology and directional transport [17]. The formation of LC phases typically requires surfactants, which act as stabilizers by adsorbing onto the surface nano- or micro- sheets [18, 19]. This prevents aggregation and restacking through steric hindrance, thereby keeping the sheets well-dispersed and mobile in aqueous solution. In particular, polyvinylpyrrolidone (PVP) also provides steric hindrance and dispersion stability, assists exfoliation, controls growth direction and shape-selective solvothermal synthesis as a nonionic surfactant [20, 21].

WS$_2$ nanosheets and reduced graphene oxide (rGO) are both widely used in sensor technologies due to their unique properties [22, 23]. WS$_2$ nanosheets have a high surface-area-to-volume ratio and excellent electronic properties, making them suitable for detecting various substances at low concentrations [22]. rGO is utilized for its low electrical noise, tunable conductivity, and ability to detect environmental changes through charge modulation or surface adsorption, contributing to highly sensitive and stable sensor platforms [23]. However, individual $WS_2$- and rGO- based sensors exhibit drawbacks that limit their practical utility. $WS_2$ suffers from low conductivity and aggregation, leading to issues like incomplete recovery and baseline drift in applications such as sensing and catalysis, while rGO suffers from moderate hysteresis and reduced selectivity in humid environments owing to its predominantly negative humidity sensitivity (resistance decrease with increasing RH. To overcome these challenges, hybrid nanostructures integrating $WS_2$ and rGO have been developed to exploit synergistic effects for enhanced performance in various applications, particularly in energy storage and catalysis. The $WS_2$/rGO hybrid nanocomposites are typically synthesized via hydrothermal methods, which effectively create a well-integrated structure [24] by combining the high adsorption capacity of $WS_2$ nanoflakes on the rGO platform [25]. Adsorption/desorption mechanisms in these structures are strongly dependent on the RH level. Different mechanisms occur when a humidity sensor is exposed to low or high RH [23]. A transition from electronic conductivity at low RH to a proton-hopping cascade (Grotthuss) mechanism becomes evident under high-humidity conditions. The transition to the Grotthuss proton transport mode should occur at lower RH in a material with LC properties, since this mode requires the formation of several ordered layers of water adsorbed on the surface. Such layers are much more easily formed in the interflake spaces of LCs.

This work presents the synthesis, characterization and optimization of $WS_2$/rGO hybrid nanocomposites by PVP elucidating the underlying mechanisms to advance the next-generation humidity sensing. Particularly, it introduces a novel approach to humidity sensing using 1T-$WS_2$/PVP/rGO liquid crystalline films, which integrate the metallic advantages of 1T-$WS_2$, the stabilizing role of PVP and the conductive framework of rGO into aligned LC structures for enhanced sensitivity, rapid response and environmental robustness.

## 2. Materials and Methods

### 2.1. Materials

PVP K17 (CAS No.: 9003-39-8), $M_w$= 7000-11000 was purchased from Ataman Chemicals. The graphite flexible foil gasket sheet of > 99 % purity (CAS No.: 7782–42-5), thiourea (CAS No.: 62-56-6), solvents and all the chemicals used in the experiments were purchased from Sigma-Aldrich Chemical Co.

### 2.2. Hydrothermal synthesis of 1T-$WS_2$/PVP/rGO hybrids

A mixture of 0.7 g of tungstic acid ($H_2WO_4$), 3 g of oxalic acid ($H_2C_2O_4$), 3 g of thiourea ($CS(NH_2)_2$) as the sulfur source, and 3 g of PVP was placed in a Teflon-lined autoclave, followed by the addition of 80 mL of an electrochemically exfoliated graphene oxide aqueous solution. The electrochemical exfoliation of GO is detailed in previously reported studies [14, 26]. A magnetic stir bar was introduced into the reactor, which was subsequently sealed in a stainless-steel shell and heated to 150 °C under continuous stirring to ensure homogenization and initiate the reaction. The mixture was maintained at this temperature for 2 hours, after which the autoclave was transferred to a preheated oven at 220 °C and held for 8 hours. Upon completion, the system was naturally cooled to room temperature, and the resulting product was collected by centrifugation, washed with deionized water and acetone, and finally dried under vacuum at 50 °C.

### 2.3. Preparation of colloidal solution of 1T-WS$_2$/PVP/rGO for LC phase formation

WS$_2$ and rGO nano- and micro- sheets are intrinsically hydrophobic, making the preparation of stable colloidal dispersions in polar solvents a significant challenge. To address this, the WS$_2$/rGO hybrid nanocomposites were synthesized in the presence of PVP, which improves the solubility in water and other polar media. By adsorbing onto the *ab*-plane, PVP not only stabilizes the dispersion but also suppresses particle growth along the *c*-axis [27]. **Table S1** in the Supplementary Information (**SI**) demonstrates detailed SEM and DLS analysis on the PVP's contribution in WS$_2$ sheets' planarization. Moreover, PVP- capped sheets demonstrate enhanced packing, facilitating the fabrication of flexible films via filtration and drop-casting. LC phase formation from a colloidal solution of GO was well described in previously reported studies [14, 28, 29]. A colloidal solution of WS$_2$/rGO/PVP was prepared in a water solution by ultrasonically treating 3 mg of powder in 1 ml of solution for one hour. **Figure S1** demonstrates the role of the LC phase in enhancing film uniformity.

### 2.4. Sample preparation and humidity testing setup

A commercially available interdigital electrode (IDE) with gold contacts was used as a substrate for the deposition of the functional thick film *via* standard drop-casted method (**Figure 1**).

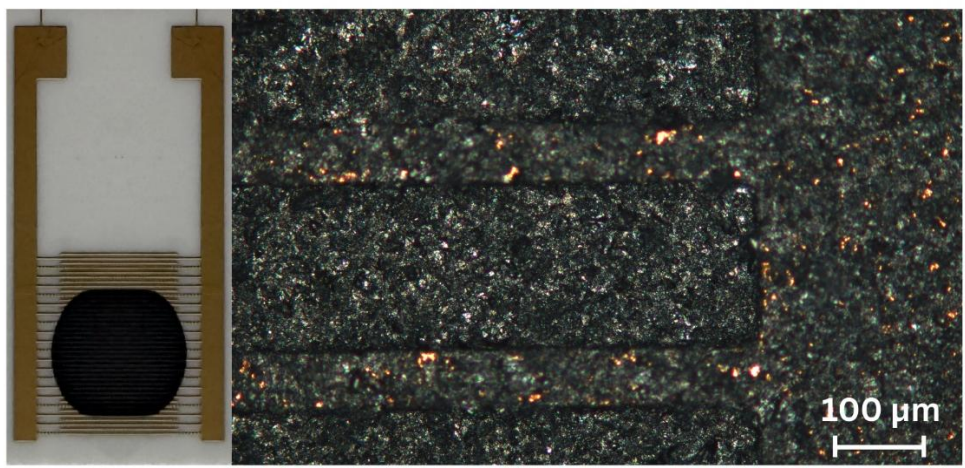

**Figure 1**. Photo and optical micrograph of the sensor.

The IDE was fabricated on an alumina (Al$_2$O$_3$) ceramic substrate with dimensions of 10 × 20 mm. The IDE structure consisted of 20 finger pairs, with a finger width of 83 µm and an inter-electrode spacing of 120.5 µm evaluated using an optical microscope (**Figure S2**). The electrodes were formed *via* standard photolithography technique by a multilayer metal stack Ti/Cu/Ni/Au, where titanium served as an adhesion layer. The nominal thicknesses of the individual layers were approximately 0.1–0.5 µm (Ti), 5 µm (Cu), 3 µm (Ni), 1 µm (Au). The synthesized WS$_2$/PVP/rGO hybrid was drop-casted onto the electrodes **Figure 2(a)**, and the device under test (DUT) was subsequently characterized using the humidity testing setup as shown in **Figure 2(b)**. Experiments were carried out using a Zennium X electrochemical workstation (Zahner-Elektrik, Germany) with an applied voltage of 0.01 V. The DUT was placed inside a sealed chamber, where humidified air was introduced. Humid air was generated by passing dry air through a gas bubbler containing saturated salt solutions, corresponding to RH values of 7%, 11%, 23%, 33 %, 43%, 75 %, 84 %, and 94 %. The chamber temperature was maintained at 23 °C and the humid air was supplied at a constant flow rate of 200 sccm. The standard exposure time to humidity was 60 seconds. Additionally, measurements were performed with exposure times of 30 and 90 seconds to study time-dependent behavior. Data acquisition was initiated after signal stabilization to ensure reliable results.

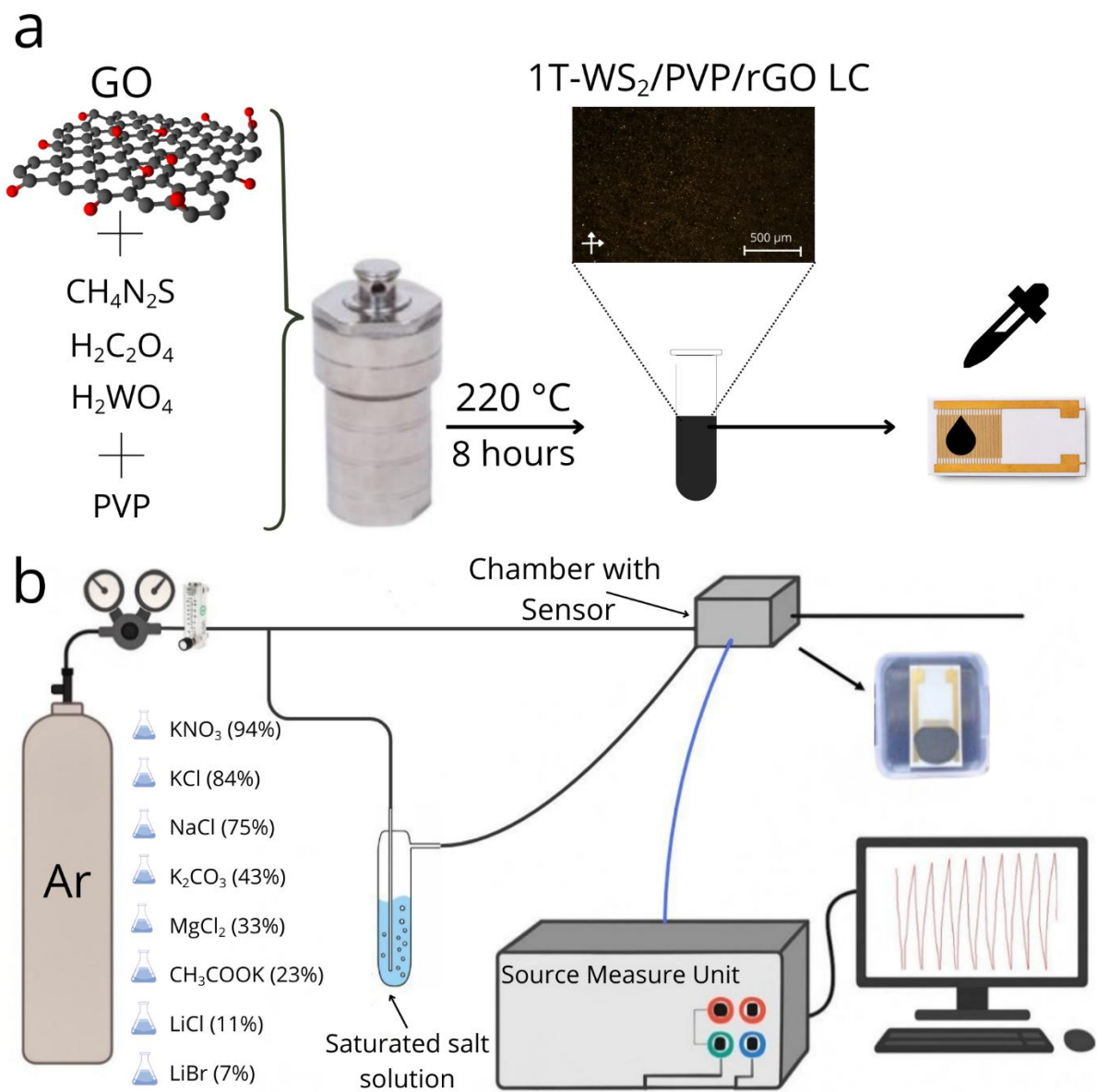

**Figure 2**. Schematic representation of WS$_2$/PVP/rGO film preparation process **(a)**, and the setup used for humidity sensing measurements **(b)**.

## 2.5. Instrumentations

Crystallographic data for the WS$_2$/PVP/rGO hybrid nanocomposite was acquired through XRD analysis (MiniFlex instrument, Rigaku, Japan). Chemical composition and valence states of the elements were studied by XPS system (Thermofisher K-Alpha, United States). The chemical structure of the synthesized compounds was analyzed using FTIR-ATR spectrometry (Nicolet 5700, Thermo Electron, United States). Bond types and hybridization were identified via AFM-combined Raman spectroscopy (LabRAM HR Evolution, HORIBA, Japan). The optical properties were characterized using UV–Vis spectrophotometry (Cary 60, Agilent, United States). The morphology and chemical composition were analyzed using scanning electron microscopy (SEM) (Prisma E from Thermo Fisher Scientific, United States), and transmission electron microscopy (TEM), operated at 200 kV and equipped with a high-annular angular dark-field detector for scanning TEM imaging (STEM-HAADF) and a Super-X detector for energy-dispersive X-ray spectroscopy elemental mapping in STEM mode (STEM-EDX) (Talos F200X G2, Thermo Fisher Scientific, United States). Thermal properties were explored by differential scanning calorimeter (STARe System TGA/DSC 3+, Mettler Toledo, Switzerland). Particle size analysis of the synthesized materials were implemented utilizing the DLS technique (Litesizer 500, Anton Paar, Austria), The LC phases of WS$_2$/PVP/rGO hybrid structure were examined using polarized optical microscopy (POM, MP920, BW Optics, China). The electrical properties were measured using the four-point probe method (Veeco Instruments, Inc., United States). Resistive behavior of the synthesized films was confirmed by electrochemical impedance spectroscopy (EIS) (photoelectrochemical workstation, Zahner-Elektrik, Germany).

## 3. Results and Discussion

### 3.1. Structural Analysis of 1T-WS$_2$/PVP/rGO Hybrids

Microscopic images and characteristic spectra of the 1T-WS$_2$/PVP/rGO hybrid are shown in **Figure 3**. SEM image of 1T-WS$_2$/PVP/rGO hybrid's morphology is shown in **Figure 3(a)**. Distinct 1T-WS$_2$ flower-like structures are visible along the edges of the rGO sheet, while hexagonal WS$_2$ platelets can be observed on the surface. TEM image of the 1T-WS$_2$ sheets is shown in **Figure 3(b)**, exhibiting typical hexagonal platelet morphology with an average lateral size of approximately 300 nm. Corresponding HRTEM analysis shown in **Figure 3(c)** reveals an interlayer spacing of 0.96 nm, which corresponds to the (001) planes of the 1T-WS$_2$ phase. This increased interlayer distance, relative to that of the 2H phase (~0.63 nm) [30], is attributed to NH$_4^+$ intercalation between the layers [31, 32]. Notably, regions exhibiting an interlayer spacing of 0.63 nm, characteristic of the 2H-WS$_2$ phase, were also observed, likely due to partial structural transformation induced by the electron beam during imaging [33]. Selected Area Electron Diffraction (SAED) patterns of the 1T-WS$_2$/rGO hybrid structure are presented in **Figure 3(d)**, showing a distinct ring pattern associated with rGO. This confirms the high crystallinity of the hybrid material. TEM image of the WS$_2$/rGO hybrid's morphology is represented in **Figure S3(a)**. The EDX elemental mapping, namely a distribution of W, S and C elements is displayed in **Figure S3(b)** and the corresponding surface areas are marked in **Figure S3(c)**. Also, the EDX spectrum and corresponding atomic percentages of the elements are displayed in **Figure S3(d)**.

**Figure 3(e)** shows the XRD patterns of 1T-WS$_2$/rGO hybrid. A broad peak at 23.4° corresponds to the (002) plane of rGO, indicating an interlayer spacing of 0.38 nm. Compared to the characteristic peak of the 2H phase of WS$_2$ at 14.4°, shown in the PDF card 01-084-1398, presented in **Figure 3(e)**, a significant shift to 9.2° is observed. The (002) planes of the 2H phase were transformed into the (001) planes of the 1T phase as a result of $NH_4^+$ intercalation [31, 32]. Previous studies have reported that ammonium-intercalated phases exhibit an interplanar spacing of 9.7 Å [31, 32, 34], consistent with this, accordingly, the spectrum shown in **Figure 3(e)** reveals an interlayer spacing of 0.96 nm (or 9.6 Å). Moreover, the simulated XRD pattern for the 1T phase of $WS_2$ agree with the previously reported data [30]. The associated diffraction planes are listed in **Table S2**.

In **Figure 3(f)**, the UV-Vis absorption spectrum of the 1T-$WS_2$/rGO hybrid composite is depicted. It exhibits a prominent peak at 271 nm, which is attributed to the π-π* electronic transition associated with the aromatic C–C bonds in the rGO structure as summarized in **Table S2**. The optical band gap of the 1T-$WS_2$/rGO hybrid, determined via Tauc plot analysis, was found to be 2.28 eV, which is likely associated with the rGO component. Literature [35] revealed that chemically reduced GO exhibits band gaps ranging from 1.8 to 2.78 eV, depending on the deoxygenation degree.

Raman spectroscopy was performed to further analyze the structure and composition of the hybrid material, as shown in **Figure 3(g)**. The characteristic peaks of rGO appear at 1328.1 cm$^{-1}$ (*D* band) and 1593.9 cm$^{-1}$ (*G* band), corresponding to the disorder-induced vibrations at the edges of graphitic domains and the in-phase vibration of the sp$^2$ carbon lattice, respectively [36, 37]. The calculated intensity ratio $I_D/I_G$ was found to be 1.46. Additional peaks attributed to rGO, namely the *2D* and *D + G* bands, are observed at 2482.0 and 2893.3 cm$^{-1}$, respectively. In the low-frequency region, five distinct peaks associated with $WS_2$ were identified. A low-intensity peak at 409.5 cm$^{-1}$ corresponds to the out-of-plane $A_{1g}$ mode of the 2H phase, attributed to strong first-order Raman scattering [30, 38]. Additionally, second-order bands at 127.6, 190.8, and 315.7 cm$^{-1}$, corresponding to $J_1$, $J_2$, and $J_3$ modes, respectively, confirm the formation of the 1T phase of $WS_2$ [39]. The peak at 260.5 cm$^{-1}$ is the $E^1_g$ band related to the octahedral coordination of W in 1T-$WS_2$ [38]. Furthermore, the absence of the $E^1_{2g}$ vibrational mode and the lack of a pronounced $A_{1g}$ peak in the Raman spectrum suggest a minimal presence of the 2H-$WS_2$ phase within the 1T-$WS_2$/rGO hybrid. This observation confirms that the 1T phase predominates in the material, consistent with previously published reports [38, 39].

The 3D surface profile of the 1T-$WS_2$/rGO hybrid structure, observed from the AFM analysis and highlighting several $WS_2$ sheets along with their measured thicknesses, is presented in **Figure S4**.

**Figure 3(h)** displays the FTIR-ATR spectra of the hybrid 1T-$WS_2$/rGO. The S-S stretching vibration band at 961 cm$^{-1}$ confirms the presence of $WS_2$ in the composite, as previously reported [40]. A strong absorption at 1091 cm$^{-1}$, along with additional peaks at 1189 cm$^{-1}$ and 1054 cm$^{-1}$, is attributed to the C-O-C stretching vibrations of epoxy and C–O vibrations of alkoxy groups [29, 41, 42]. These signals suggest a significant presence of C–O–C groups located at the edges and basal planes of rGO sheets, aligning well with previously reported observations in pure rGO films [14]. Furthermore, the absorption bands observed at 838 cm$^{-1}$ and 1472 cm$^{-1}$ correspond to C–C bonds and C–H bonds of -$CH_2$ groups, which are characteristic of both rGO and PVP [29, 43-45]. The band at 1324 cm$^{-1}$ is ascribed to C–N stretching vibrations, likely arising from PVP's structure [44, 45]. The absorption band at 1434 cm$^{-1}$ is attributed to the vibrational mode of intercalated $NH_4^+$ ions, indicating their presence within the structure [46].

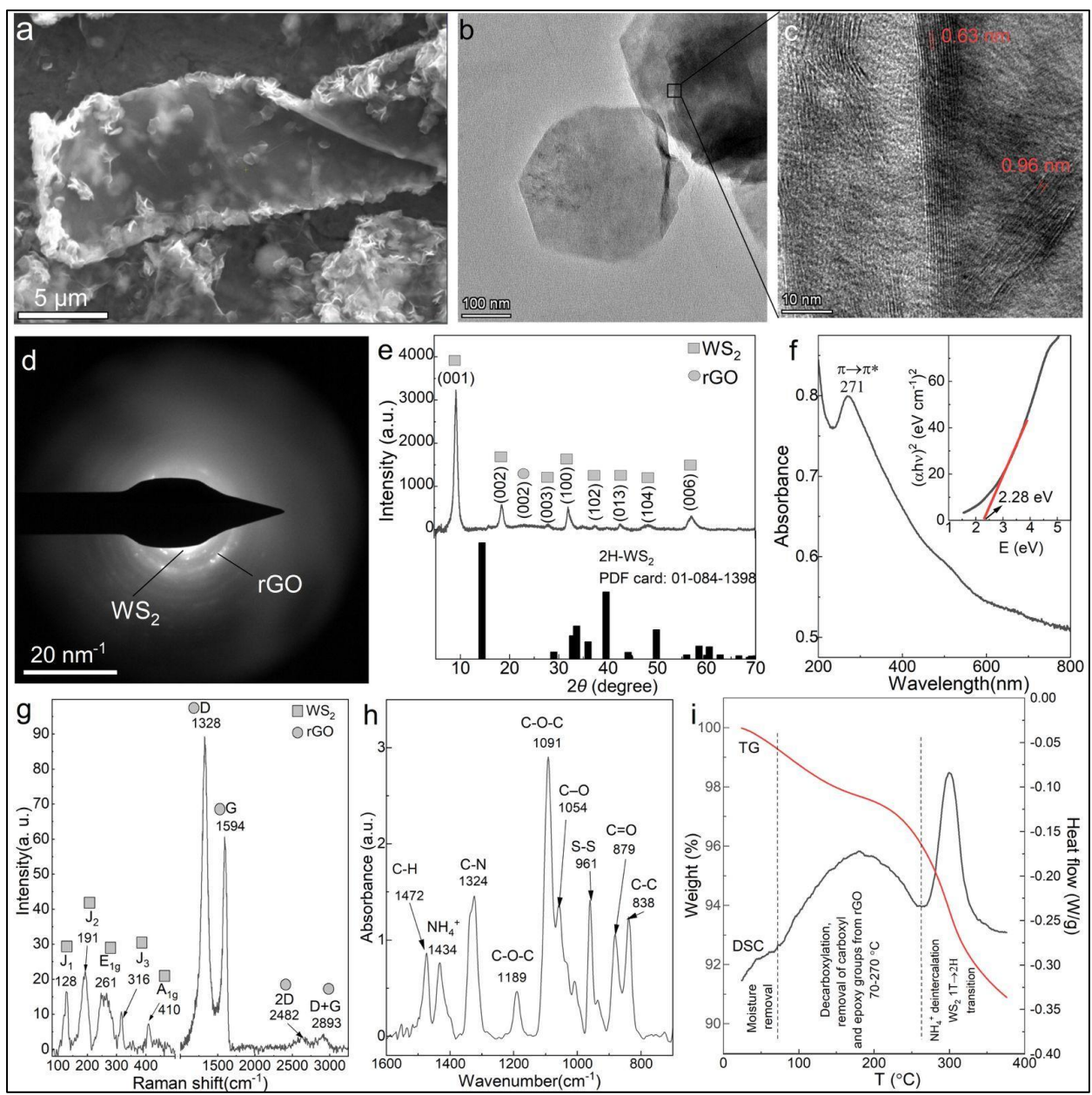

**Figure 3**. SEM image of 1T-WS$_2$/PVP/rGO hybrid nanocomposite **(a)**. TEM image of pristine WS$_2$ sheet **(b)**; HRTEM analysis **(c)** and SAED diffraction pattern **(d).** XRD spectrum and PDF card 01-084-1398. Cu$K\alpha$-radiation and scanning range: -3-145°(2θ) **(e)**. UV–Vis absorption spectrum with corresponding Tauc plot in the inset, with spectral range: 190 - 1100 nm **(f)**. The Raman spectrum of the hybrid structure at wavelength range: 200-1100 nm, with light source: 633 nm, and a 600 gr/mm grating **(g)**. FTIR-ATR spectrum with spectral range: 1600 to 600 cm$^{-1}$ using a ZnSe crystal with Happ–Genzel apodization, (ATR distortion was corrected, number of scans 32, resolution 4 cm$^{-1}$) **(h)**. DSC-TGA study of the 1T-WS$_2$/PVP/rGO in argon gas flow at T = 400 °C and V = 5°/min **(i).**

To analyze the thermal stability of the hydrothermally prepared 1T-$WS_2$/PVP/rGO, the phase was subjected to the DSC - TGA study in an argon flow (gas flow rate of 80 mL/min) at a heating rate of 5°/min from room temperature to 400 °C as shown in **Figure 3(i)**. Two main exothermic events accompanied by mass loss are observed on the DSC curve in the temperature intervals of 80-260°C and 270-400 °C, respectively. The first mass loss was the result of decarboxylation and removal of residual carboxyl and epoxy groups from the rGO sheets [47, 48]. At the temperature range 270-370 °C, a weight loss of about 4 % from the initial mass was observed in the TGA curve. The subsequent exothermic transition was detected on the DSC curve, beginning at approximately 270 °C. It is well-established that annealing at approximately this temperature range results in the complete removal of ammonia ions and promotes the phase transformation of $WS_2$ from the metastable 1T phase to the thermodynamically stable 2H phase [32, 49]. The thermal effect associated with this transition, evaluated from the DSC peak area using the instrument software, was 56.1 kJ/g. The XRD analysis of the heating product confirms 2H phase formation at 370 °C.

XPS was employed to further investigate the chemical composition and valence states of the elements in the 1T-$WS_2$/PVP/rGO hybrid. For comparison, 1T-$WS_2$/PVP and rGO/PVP reference samples were additionally synthesized following the same procedure described in Section 2.2 (Hydrothermal synthesis of the 1T-$WS_2$/PVP/rGO hybrid). The full survey XPS spectra of 1T-WS2/PVP, rGO/PVP, and 1T-$WS_2$/PVP/rGO are shown in **Figure S5** and confirm the presence of W, S, C, O, and N elements in the samples. **Figure 4** presents the high-resolution deconvoluted XPS spectra of the W 4f, S 2p, C 1s, O 1s, and N 1s core levels, while the corresponding peak positions and assignments are summarized in **Table 1.**

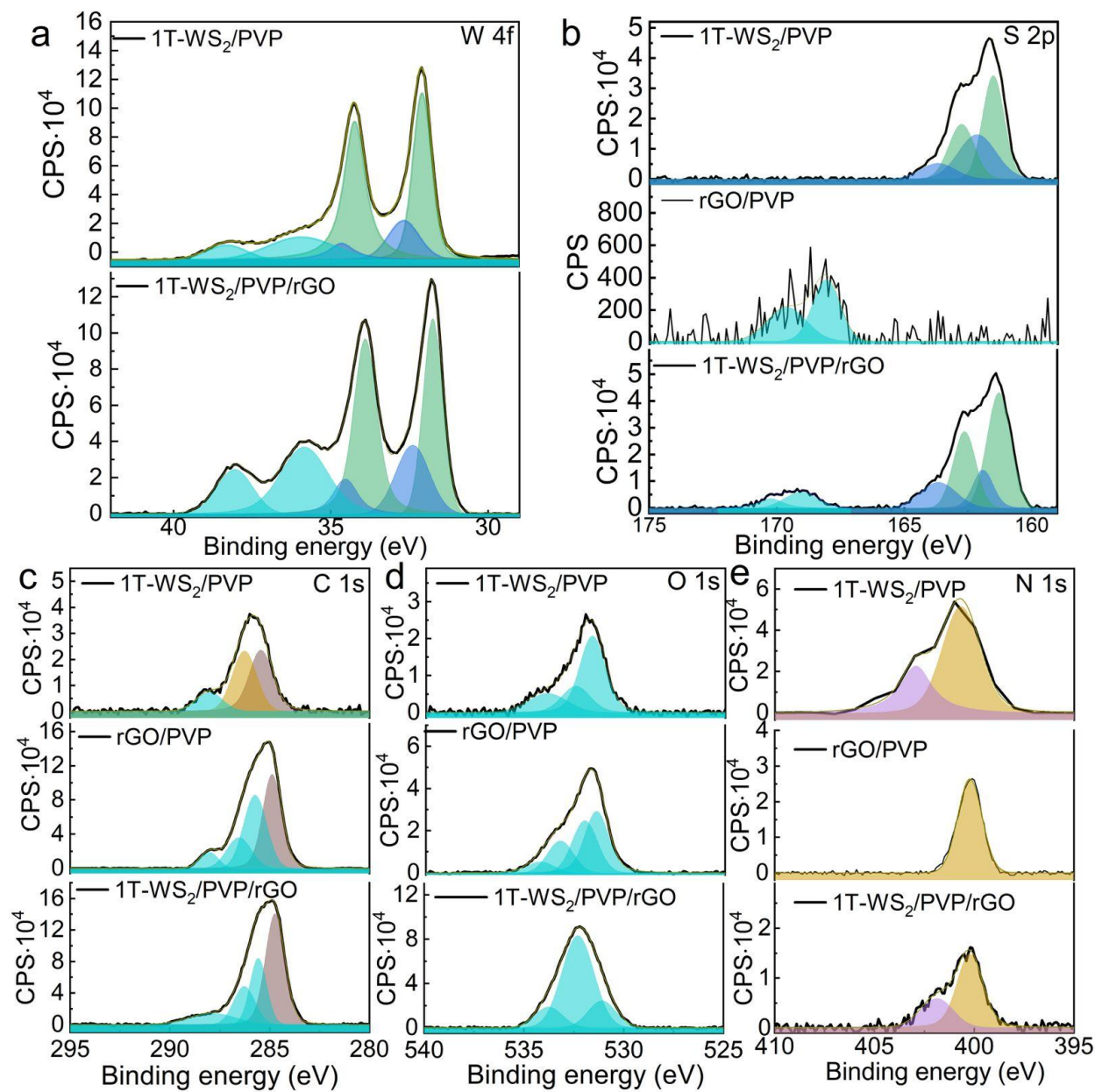

**Figure 4**. XPS analysis of 1T-WS$_2$/PVP/rGO in comparison with 1T-WS$_2$/PVP and rGO/PVP. High-resolution deconvoluted spectra of W 4f (a) for 1T-WS$_2$/PVP and 1T-WS$_2$/PVP/rGO, S 2p (b), C 1s (c), O 1s (d), and N 1s (e) for 1T-WS$_2$/PVP, rGO/PVP, and the 1T-WS$_2$/PVP/rGO hybrid. The fitted components correspond to different chemical contributions: 1T phase of WS$_2$ (blue), 2H phase of WS$_2$ (green), and oxygen-related bonds (cyan). In the C 1s spectra, C–C bonds are shown in brown, C–N bonds in yellow, and oxygen-containing carbon species in cyan. In the N 1s spectra, C-N is shown in yellow, and ammonium species are represented in purple.

The XPS analysis of both 1T-WS$_2$/PVP and 1T-WS$_2$/PVP/rGO reveals a dominant W 4f doublet characteristic of the metallic 1T-W(IV) phase, accompanied by a minor contribution from the semiconducting 2H phase. The spin–orbit splitting of ~2.1 eV observed for the W 4f core level in both systems is consistent with tungsten in the $W^{4+}$ oxidation state. Quantitative analysis of the fitted W 4f components indicates that the fraction of the 1T phase is approximately 79.4% in 1T-$WS_2$/PVP and 71.7% in 1T-$WS_2$/PVP/rGO (**Figure 4(a)**). Compared to 1T-$WS_2$/PVP, the W 4f binding energies in the 1T-$WS_2$/PVP/rGO hybrid are shifted toward lower values, indicating enhanced electronic screening and interfacial charge redistribution. In addition, small contributions from oxygenated $WO_x$ species are detected, most likely corresponding to surface tungsten oxides (e.g., $WO_3$), originating from post-synthesis surface oxidation, as commonly reported for $WS_2$-based materials [50, 51].

A similar trend is observed in the high-resolution S 2p spectra (**Figure 4(b)**). For rGO/PVP, the S 2p region exhibits a very weak signal, which is attributed to sulfate species likely originating from residual sulfur-containing groups in the initial GO precursor. In contrast, the S 2p spectrum of 1T-$WS_2$/PVP/rGO displays well-defined sulfide-related peaks with a noticeable shift toward higher binding energies, indicating modifications in the chemical environment of sulfur due to interfacial interactions within the hybrid.

XPS analysis reveals a systematic shift of the W 4f and S 2p ($S^{2-}$) core levels toward lower binding energies in the 1T-$WS_2$/rGO/PVP hybrid compared to 1T-$WS_2$/PVP. The observed changes in binding energy indicate the presence of charge transfer in the 1T-$WS_2$/PVP/rGO hybrid at the interface between rGO and 1T-$WS_2$. Considering the p-type conductivity of rGO, this behavior can be attributed to interfacial charge redistribution driven by Fermi level alignment between rGO and metallic 1T-$WS_2$. In this process, hole transfer from rGO to $WS_2$, or equivalently, electron accumulation in $WS_2$, enhances electronic screening around the W and S atoms, resulting in reduced core-level binding energies. In contrast, the sulfate-related S 2p components exhibit higher binding energies in rGO/PVP, indicating weaker electronic screening in the absence of $WS_2$. The reduced binding energies of the sulfate species observed in the composite further support the presence of interfacial charge redistribution rather than chemical transformation, confirming strong electronic coupling between rGO and $WS_2$ [12].

As shown in **Figure 4(c)**, the C 1s XPS spectra reveal the presence of carbon atoms in different functional groups for all investigated samples. The C 1s spectrum of 1T-$WS_2$/PVP exhibits characteristic contributions from the PVP polymer, including C–C, C–N, and C=O bonds, in good agreement with previous reports [52, 53] The rGO/PVP sample displays additional oxygen-containing functional groups, such as C–O–C, C–OH, and C=O, together with the C–C component, resulting in an O/C atomic ratio of approximately 0.50. In the 1T-$WS_2$/PVP/rGO composite, the same functional groups are observed; however, the O/C ratio decreases to ~0.43, indicating a greater reduction of oxygen-containing groups. In the C 1s spectra of rGO/PVP and 1T-$WS_2$/PVP/rGO, the peak corresponding to C–N bonds cannot be clearly distinguished because its binding energy is very close to that of C–O–C groups, while the epoxy component naturally diminishes in the hybrid system.

The O 1s high-resolution XPS spectra provide further insight into the oxygen chemical states and interfacial interactions in the studied materials. For the 1T-$WS_2$/PVP sample, the O 1s spectrum consists of contributions associated with residual tungsten oxide species ($WO_3$), as well as oxygen-containing groups originating from the PVP polymer, including C=O and C–O–C bonds (**Figure 4(d)**). The presence of the C–O–C component may be related to partial oxidation of the polymer chains under the synthesis conditions.

In the rGO/PVP sample, the O 1s spectrum is more complex and includes components assigned to O–C=O, C=O, C–OH, and C–O–C functional groups, reflecting the oxygen-rich surface chemistry of reduced graphene oxide. In contrast, the O 1s spectrum of the 1T-WS$_2$/PVP/rGO composite is dominated by the C=O component, indicating a reduced contribution from hydroxyl groups.

The N 1s XPS spectra of all three materials exhibit a dominant peak corresponding to C–N bonds, which originates from the pyrrolidone groups of PVP (**Figure 4(e)**). In addition, nitrogen species attributed to NH$^{4+}$ are observed exclusively in the WS$_2$-containing samples (1T-WS$_2$/PVP and 1T-WS$_2$/PVP/rGO). The presence of NH4+ is associated with its participation in the stabilization and formation of the metallic 1T phase of WS2 during synthesis, in agreement with previous reports [31, 32].

**Table 1.** Binding energies of the fitted XPS peaks for 1T-WS$_2$/PVP, rGO/PVP, and 1T-WS$_2$/rGO/PVP, along with their corresponding assignments.

| | | Binding Energy (eV) | | | | | |
|---|---|---|---|---|---|---|---|
| 1T-WS$_2$/PVP | W 4f | 32.1 | 32.6 | 34.2 | 34.7 | 35.9 | 38.3 |
| rGO/PVP | | – | – | – | – | – | – |
| 1T-WS$_2$/PVP/rGO | | 31.8 | 32.4 | 33.9 | 34.5 | 35.8 | 38.1 |
| Assignment | | 1T-W (IV) 4f$_{7/2}$ | 2H-W (IV) 4f$_{7/2}$ | 1T-W (IV) 4f$_{5/2}$ | 2H-W (IV) 4f$_{5/2}$ | W (VI) 4f$_{7/2}$ | W (VI) 4f$_{5/2}$ |
| 1T-WS$_2$/PVP | S 2p | 161.5 | 162.2 | 162.8 | 163.7 | – | – |
| rGO/PVP | | – | – | – | – | 168.1 | 169.6 |
| 1T-WS$_2$/PVP/rGO | | 161.3 | 161.9 | 162.6 | 163.7 | 169.0 | 170.2 |
| Assignment | | 1T-S 2p$_{3/2}$ | 2H-S 2p$_{3/2}$ | 1T-S 2p$_{1/2}$ | 2H-S 2p$_{3/2}$ | S (VI) 2p$_{1/2}$ | S (VI) 2p$_{3/2}$ |
| 1T-WS$_2$/PVP | C 1s | 285.5 | – | – | 286.3 | 288.0 | |
| rGO/PVP | | 284.9 | 285.7 | 286.5 | – | 288.0 | |
| 1T-WS$_2$/PVP/rGO | | 284.7 | 285.6 | 286.3 | – | 287.9 | |
| Assignment | | C-H; C-C | C-OH | C-O-C | C-N | C=O | |
| 1T-WS$_2$/PVP | O 1s | 531.7 | – | 532.5 | – | 533.9 | |

| | | | | | | |
|---|---|---|---|---|---|---|
| rGO/PVP | | – | 531.4 | 532.0 | 533.3 | 534.2 |
| 1T-WS$_2$/PVP/rGO | | – | 531.1 | 532.3 | – | 533.8 |
| Assignment | | WO$_3$ | O-C=O | C=O | C-OH | C-O-C |
| 1T-WS$_2$/PVP | N 1s | 400.7 | | | 402.9 | |
| rGO/PVP | | 400.2 | | | – | |
| 1T-WS$_2$/PVP/rGO | | 400.2 | | | 401.9 | |
| Assignment | | C-N | | | NH$_4^+$ | |

### 3.2. Humidity sensing of the 1T-WS$_2$/PVP/rGO hybrid

To analyze the response of the DUT, the humidity sensing experiments were conducted using different humidity conditions by the assembled setup shown in **Figure 2**. This was achieved by bubbling argon through a series of saturated salt solutions, each providing a specific water vapor concentration in the chamber atmosphere where the DUT is placed. As humidity sensors can operate based on different transduction principles [54-58], the resistive behavior of the DUT using electrochemical impedance spectroscopy (EIS) was confirmed (**Figure 5(a)**). The phase plots in dry, 7 % and 94 % humid air indicate no significant phase shift (phase ≈ 0° in the range of 0.1 to $10^4$ Hz), with a slight increase observed at higher frequencies ($10^4$ - $10^5$ Hz), most likely due to parasitic effects. The magnitude plots indicate that impedance Z remains nearly constant across the entire AC frequency range, as evidenced by the Bode plot, where the magnitude traces appear as horizontal lines parallel to the frequency axis. These results confirm that the DUT exhibits predominantly resistive behavior. The interpretation of the results was done based on previously reported data [59]. Besides, the sheet resistance of the WS$_2$/PVP/rGO was also measured by the four-point probe method which indicates 45 Ω average value. **Figure 5(b)** represents the current-voltage characteristics for the 1T-WS$_2$/PVP/rGO hybrid film on the IDE for dry air and eight RHs (7 %, 11 %, 23 %, 33 %, 43 %, 75 %, 84 %, and 94 %). The I(V) curves clearly demonstrate that the rate of current change increases significantly as humidity decreases. The dynamic response-recovery curves at the above-mentioned RHs are displayed in **Figure S6(a-c)**. As one can see, the response demonstrates excellent repeatability and it can be defined by **Eq. (1)**:

$$\frac{(I_h - I_0)}{I_0} \times 100\% \qquad (1)$$

where $I_h$ is current in different testing RHs, and $I_0$ is current in 0 % RH. Normalized DUT response to various RH levels clearly indicates a decrease in current in response to rising humidity in the carrier gas, caused by water adsorption on its surface or within its volume (**Figure 5(c)**). Fitted current versus RH is shown in **Figure 5(d)**. A nonlinear correlation is modeled using an exponential fitting curve. An exponential dependence of the resistance on RH was previously reported in [60]. The ΔI(RH) dependence reveals the opposite trend, namely the change in the peak amplitude of current increases with increasing relative humidity. This is most likely attributed to the increased conformational changes in PVP and the corresponding reduction in the order parameter of the 1T-WS$_2$/PVP/rGO LC structure at higher RH levels. The response/recovery times taken from 10 % to 90 % of the peak values was calculated for eight peaks (**Table S3**) and then averaged as seen in **Figure 5(e)**. In **Figure 5(f)**, the response rate on the RH was presented, considering smooth and jagged sections observed in the I(t) dependence shown in **Figure (f)**. Particularly, the humidity was applied for a total of 60 seconds, and the response rate coefficient was calculated within the 10–50 second interval, where the response curve exhibits a linear region. Each point represents the average of eight response curves for each humidity level. Detailed calculation results are tabulated in Table S3. Finally, for intermediate 33 % RH the experiments were repeated for various time intervals, given that the DUT's sensitivity is highly time dependent as displayed in **Figure 5(g)** [61].

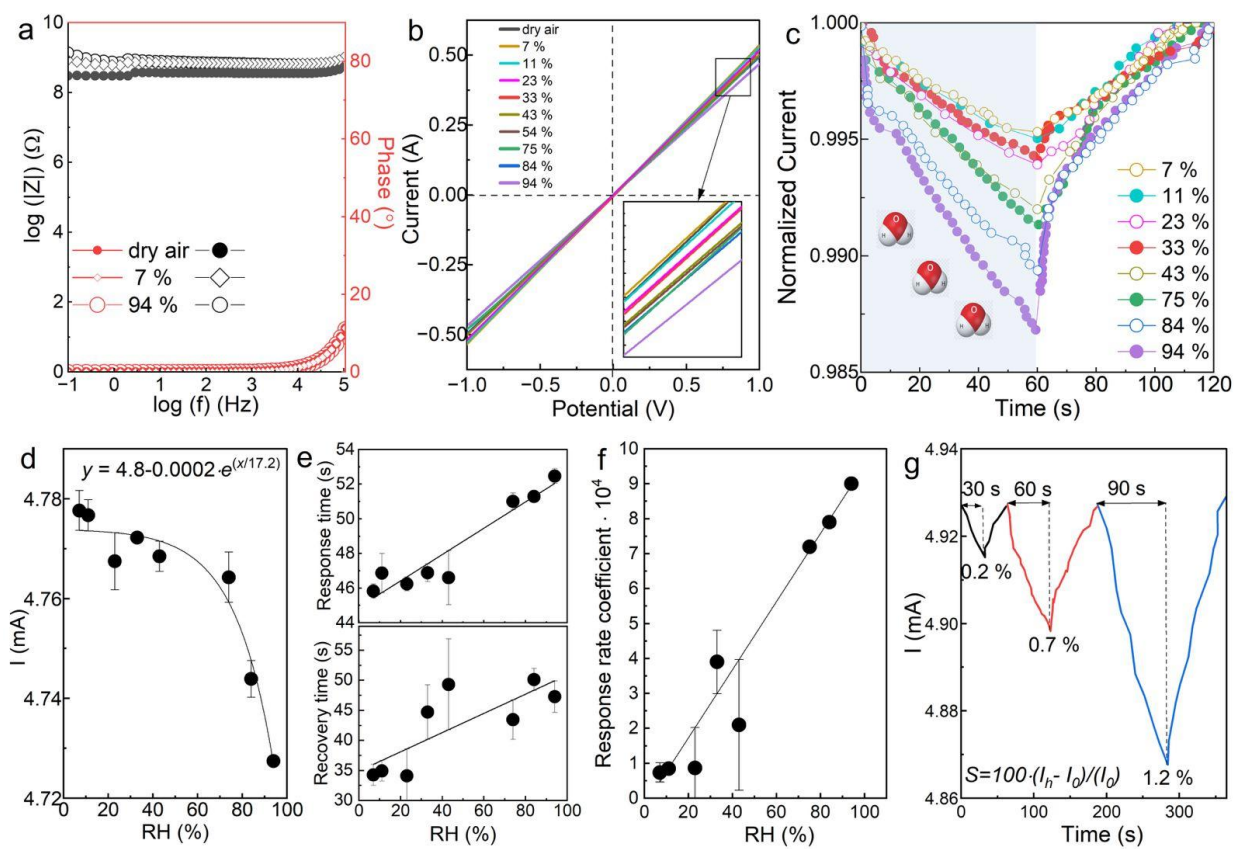

**Figure 5**. Humidity sensing properties of the 1T-WS$_2$/PVP/rGO film. Bode plot of humidity sensor 1T-WS$_2$/PVP/rGO in dry air, at 7 % and 94 % RH levels **(a)**. The black curve represents the magnitude plot, and the red curve represents the phase plot. Current-voltage I(V) characteristics for dry air and all RH levels. Inset: zoomed view between 0.8 V and 0.9 V, highlighting current trend **(b)**. Normalized DUT response to various RH levels **(c)**. The blue region indicates the period of humidity exposure. Exponential dependence of current on RH levels **(d)**. Sensor response/recovery times at different RH levels **(e)**. Dependence of the response rate coefficient on RH **(f)**. Effect of humidity exposure time on sensor sensitivity for intermediate 33 % RH level **(g)**. The switching between modes was controlled using a plumbing T-connector allowing for easy

switching between dry and humid gas flow. All measurements were performed at room temperature 23°C.

At present, a wide range of humidity measurement devices, constructed from various materials and based on diverse operating principles, are commercially available and under active development. These sensors can be classified in several ways which covers key humidity-sensitive materials, the underlying sensing mechanisms, the primary performance characteristics, and the effects of temperature on their operation and accuracy. While the performance comparison is important, their direct comparison is not relevant since it depends on many factors, such as humidity exposure time, MUT thickness, calculation method, etc. In Supporting Information of [62], Table S1 summarizes sensing performance of the reported humidity sensors based on 2D materials. Sensing performance of the humidity sensor calculated by different equations [62] is presented in Table S4.

Additionally, we have done proof-of-concept simple demonstrations of real-world humidity sensing, beyond measurements in a controlled chamber, where the response time is 3s and 4.7 s (**Figure S7**). The faster response/recovery (<10 s) depends on the flow rate. In our measurements the humid air was supplied at a constant flow rate of 200 sccm, while usually in similar works it is 1000 sccm [63-65].

Schematic illustration of the humidity sensing platform and its operating mechanism under low and high RH conditions is presented in **Figure 6**.

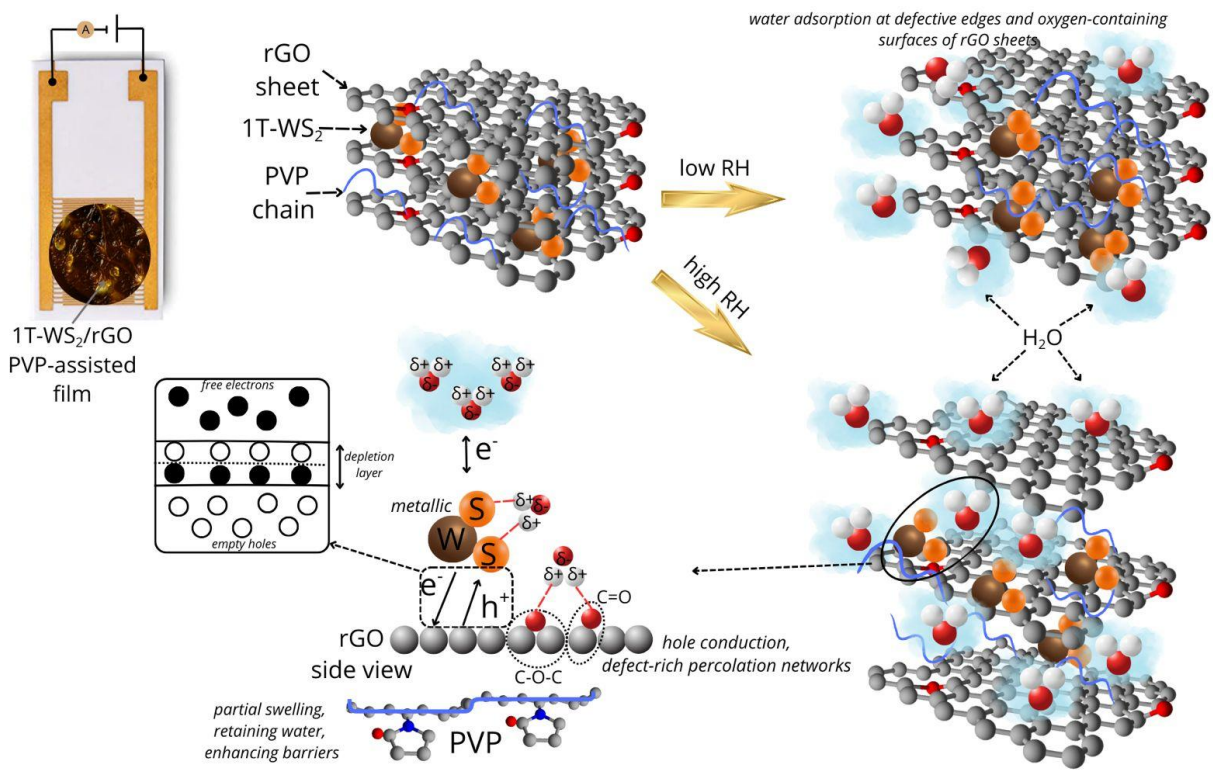

**Figure 6.** Schematic illustration of the DUT and its operating mechanism under low and high RH levels.

Typically, resistive-type humidity sensors exhibit negative humidity sensitivity behavior, characterized by a decrease in resistance with rising humidity. Conversely, reports on sensors displaying a positive humidity response remain rare in the literature [25]. Positive humidity sensitivity in the proposed system might be attributed to various factors, including the intrinsic metallic properties of 1T-WS$_2$ [7], charge transfer mechanisms, and interfacial interactions modified by PVP and rGO [66]. Particularly, a depletion of carriers in 1T-WS$_2$, where water molecules could act as electron donors, leading to a decrease in the number of charge carriers and thus an increase in resistance [67]. Besides, the interaction of water molecules with the surface of 1T-WS$_2$ may cause surface passivation or modify its electronic band structure, thereby impeding rather than promoting electron transport. On the other hand, PVP is a polymer characterized by pronounced hydrophilic properties [68]. It affects the interfacial coupling between 1T-WS$_2$ and rGO. Water-induced swelling or variations in inter-flake distances caused by PVP could interrupt charge transport pathways, thereby increasing resistance [69].

The XPS-derived binding energy shifts also provide important insight into the humidity adsorption mechanism of the 1T-WS2/rGO/PVP composite. The observed decrease in the W 4f and S 2p (S$^{2-}$) binding energies indicate enhanced electronic screening caused by interfacial charge redistribution between p-type rGO and metallic 1T-WS$_2$. This charge redistribution increases the surface polarizability of WS$_2$ and strengthens local electric fields at defect sites and edges, which act as preferential adsorption centers for water molecules.

The structural stability of the 1T-WS$_2$/rGO film was confirmed by Raman mapping of the DUT after experimental treatment as shown in **Figure S8**. Estimated $I_D/I_G$ ratio after measurements was 1.7 indicating defect increase in rGO. Moreover, stability tests conducted over several weeks confirmed that the DUT maintained a nearly constant response for more than six months.

## 4. Conclusion

In this study, the hydrothermal synthesis, comprehensive characterization, and humidity-sensing performance of the 1T-WS$_2$/PVP/rGO hybrid nanocomposites were examined under varying RH levels (7%-94%). The sensing mechanism along with key performance characteristics, such as response and recovery time is thoroughly discussed. Particularly, the response and recovery times for the intermediate 33 % RH were 46.9 s and 44.7 s, respectively, measured over a 60 s interval, which yielded a sensitivity of 0.7 % at a constant humid air flow rate of 200 sccm. However, these values are strongly dependent on the time interval. The present measurements reveal that for intervals of 30 s and 90 s, the sensitivities are 0.2% and 1.2%, respectively. Importantly, the hybrid nanocomposite system exhibited positive humidity sensitivity. Unlike disposable sensors, the proposed hybrid material is reusable as it can be readily cleaned, dissolved and drop-casted on the IDE multiple times. Overall, leveraging LC alignment, the fabricated films effectively address the limitations of conventional 2D material-based sensors enabling high-performance detection for wearable and environmental applications.

# Supplementary Materials for

# Polyvinylpyrrolidone Planarized Liquid Crystalline 1T-WS$_2$/rGO Hybrid Nanocomposites-based Humidity Sensing Platform

1. Preparation of colloidal solution of 1T-WS$_2$/PVP/rGO for LC phase formation

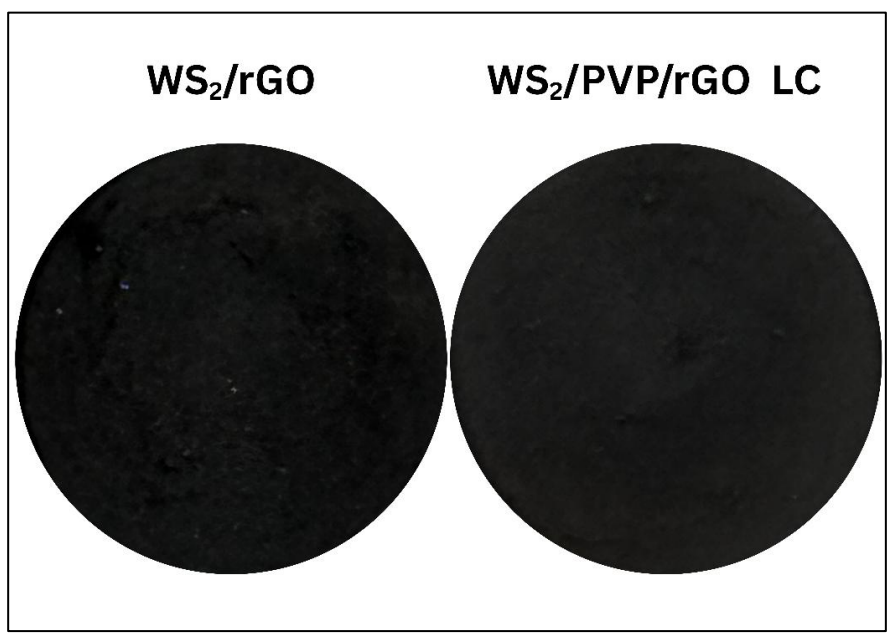

**Figure S1**. Films produced from LC **(a)** and non-LC **(b)** dispersion of solutions.

**Table S1**. Estimation of the hydrothermally synthesized pristine WS$_2$ and WS$_2$/PVP plates' sizes on the rGO platform based on SEM image and particle size distribution in a water solution (averaged over 50 WS$_2$ petals).

| SEM | | DLS | | |
|---|---|---|---|---|
| WS$_2$ | WS$_2$/PVP | WS$_2$ | WS$_2$/PVP | WS$_2$/PVP/rGO |
| 147 nm | 380 nm | 327 nm | 498 nm | 660 nm |

The differences in the sizes based on these two methods lies in the fact that SEM imaging shows the apparent lateral size of visible petals on the surface. It may miss thickness contributions or hidden parts if flakes overlap or are partially embedded. Besides, dynamic light scattering (DLS) measures the hydrodynamic diameter of particles dispersed in liquid. This includes not just the core flake but also the surrounding solvation shell such as water molecules, ions, surfactant, in this case PVP layers. In addition, possible aggregation or stacking of irregularly shaped flakes in suspension which are interpreted as equivalent spheres can also cause larger values for DLS rather than SEM measurements of the same material. Overall, it is evident that PVP plays a key role in promoting the growth of larger lateral $WS_2$ petals.

## 2. Sample preparation and humidity testing setup

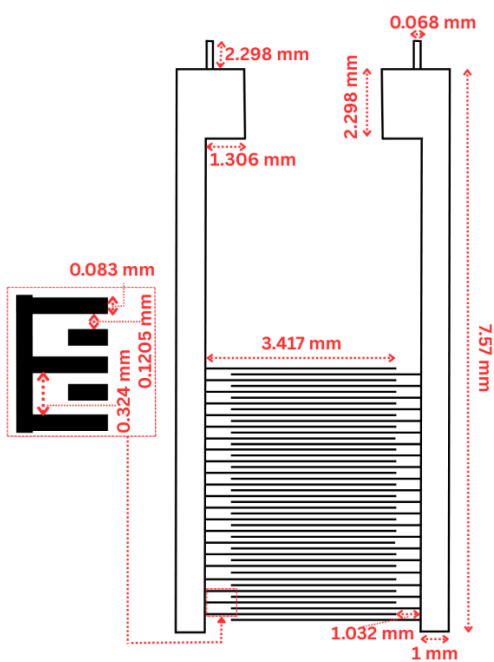

**Figure S2**. Sizes of the interdigital electrode used in the measurements.

## 3. Structural Analysis of 1T-WS$_2$/PVP/rGO hybrids

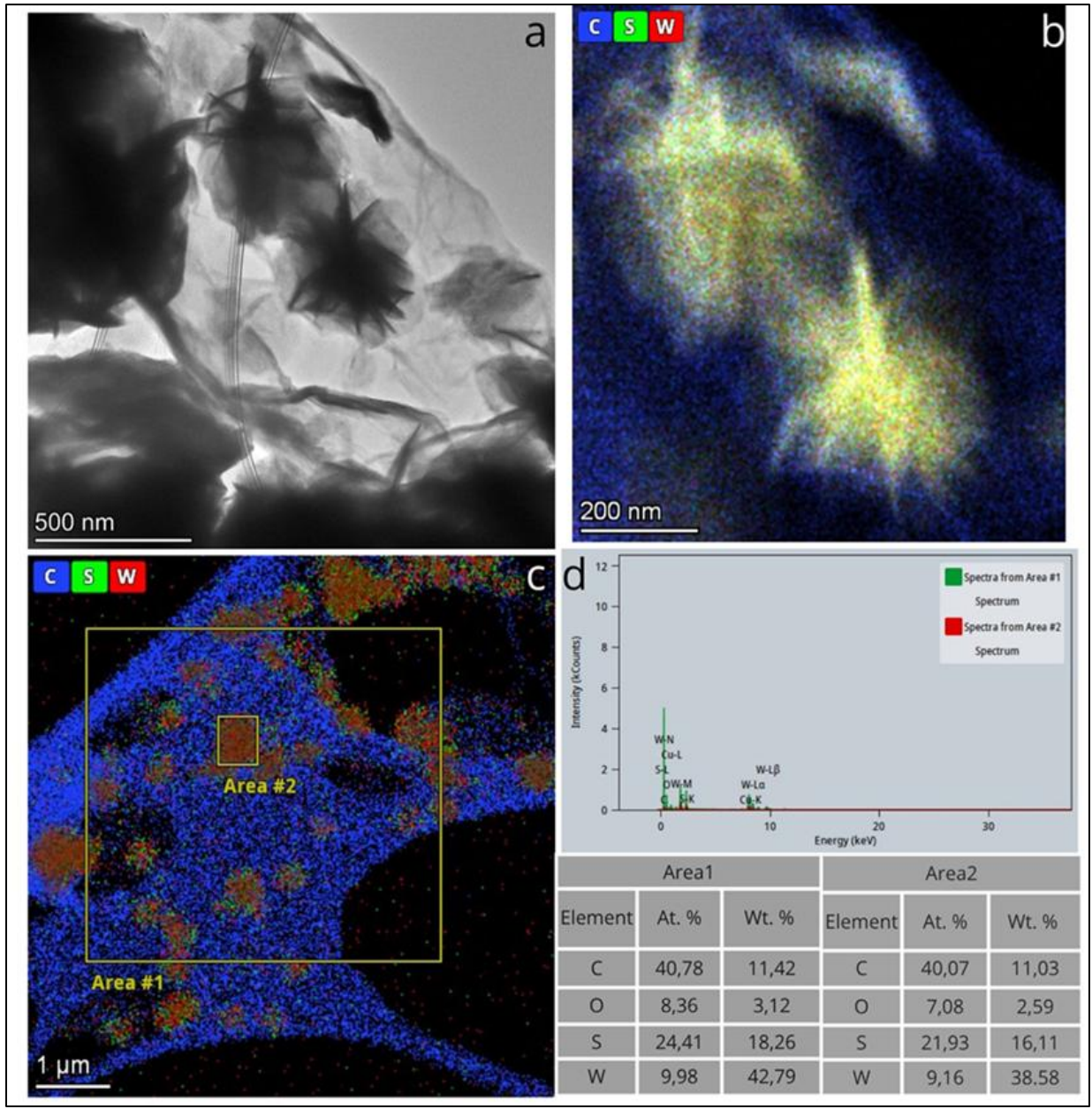

**Figure S3**. TEM image of the 1T-WS$_2$/rGO hybrid structure **(a)**. STEM-EDX chemical composition map of 1T-WS$_2$ nanoflowers on the rGO basal plane with the following color mapping: blue — carbon, red — tungsten, green — sulfur **(b)** from the corresponding surface areas **(c)**. EDX spectrum with a table showing the atomic percentage distribution of elements from the selected surfaces **(d)**.

**Table S2**. Characterization 1T-WS$_2$/rGO by XRD, UV-Vis, Raman and FTIR-ATR spectroscopy analysis.

| XRD | | |
|---|---|---|
| 2$\theta$, degree | *hkl* | Material |
| 9.2 | (001) | 1T WS$_2$ |
| 18.5 | (002) | 1T WS$_2$ |
| 23.4 | (002) | rGO |
| 27.9 | (003) | 1T WS$_2$ |
| 32.1 | (100) | 1T WS$_2$ |
| 37.5 | (102) | 1T WS$_2$ |
| 42.7 | (013) | 1T WS$_2$ |
| 48.3 | (104) | 1T WS$_2$ |
| 57.1 | (006) | 1T WS$_2$ |
| **UV-Vis** | | |
| Wavelength, nm | Transition | Material |
| 271.0 | π-π* | rGO |
| **Raman analysis** | | |
| Raman shift, cm$^{-1}$ | Vibrational modes | Material |
| 127.6 | J$_1$ | WS$_2$ |
| 190.8 | J$_2$ | |
| 260.5 | E$_{1g}$ | |
| 315.7 | J$_3$ | |
| 409.5 | A$_{1g}$ | |
| 1328.4 | D band | rGO |
| 1593.9 | G-band | |
| 2482.0 | 2D band | |
| 2893.3 | D+G | |
| **FTIR-ATR** | | |

| Wavenumber, cm$^{-1}$ | Functional group | Material |
|---|---|---|
| 1054 | C–O | rGO |
| 1091 | C–O–C | |
| 1189 | C–O–C | |
| 838 | C–C | PVP and rGO |
| 1472 | C–H | |
| 879 | C=O | PVP |
| 1324 | C–N | |
| 1434 | | $NH_4^+$ |
| 961 | S-S | $WS_2$ |

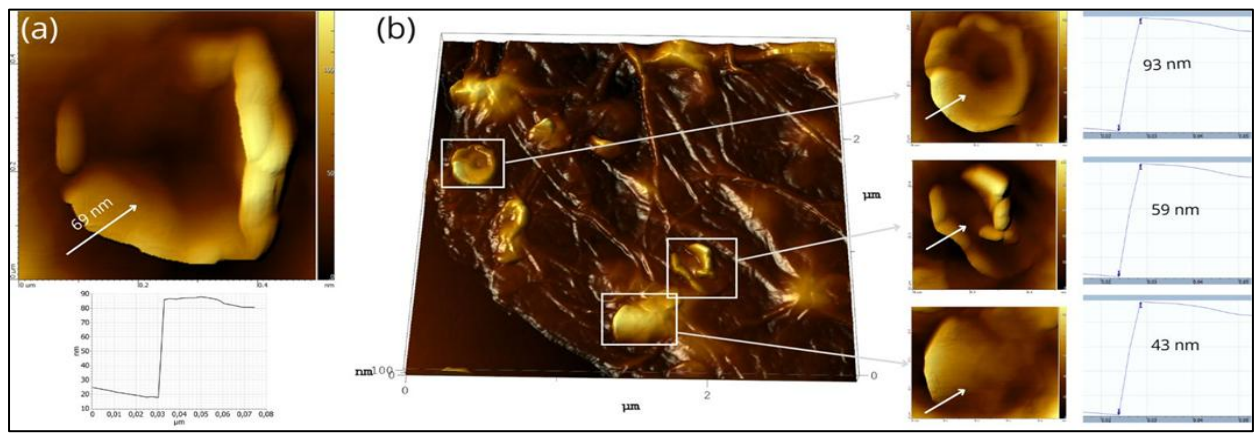

**Figure S4**. The AFM image of the WS$_2$ sheet with its indicated thickness **(a)**. The 3D surface profile of the 1T-WS$_2$/rGO hybrid structure showing several WS$_2$ sheets with their measured thicknesses **(b)**.

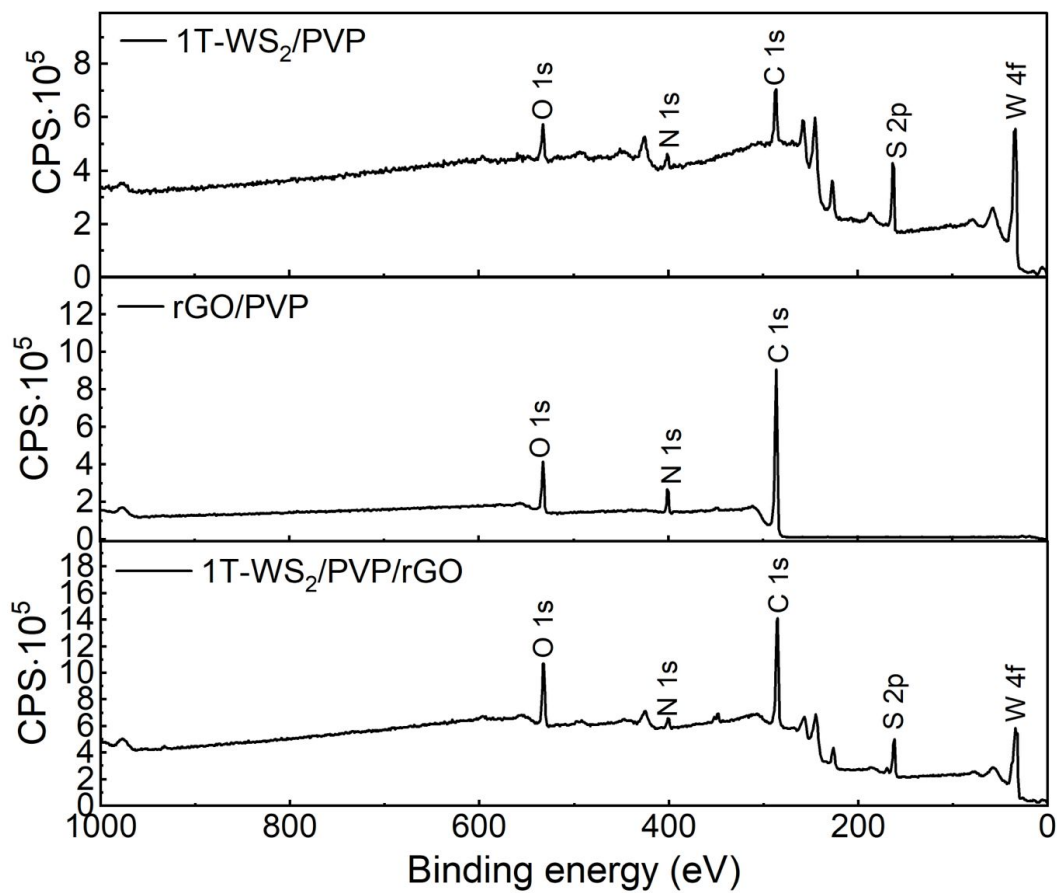

**Figure S5.** XPS survey spectra of 1T-WS$_2$/PVP, rGO/PVP, and 1T-WS$_2$/PVP/rGO, showing the positions of W 4f, S 2p, C 1s, O 1s, and N 1s core levels.

## 4. Humidity sensing of the 1T-WS$_2$/PVP/rGO hybrid

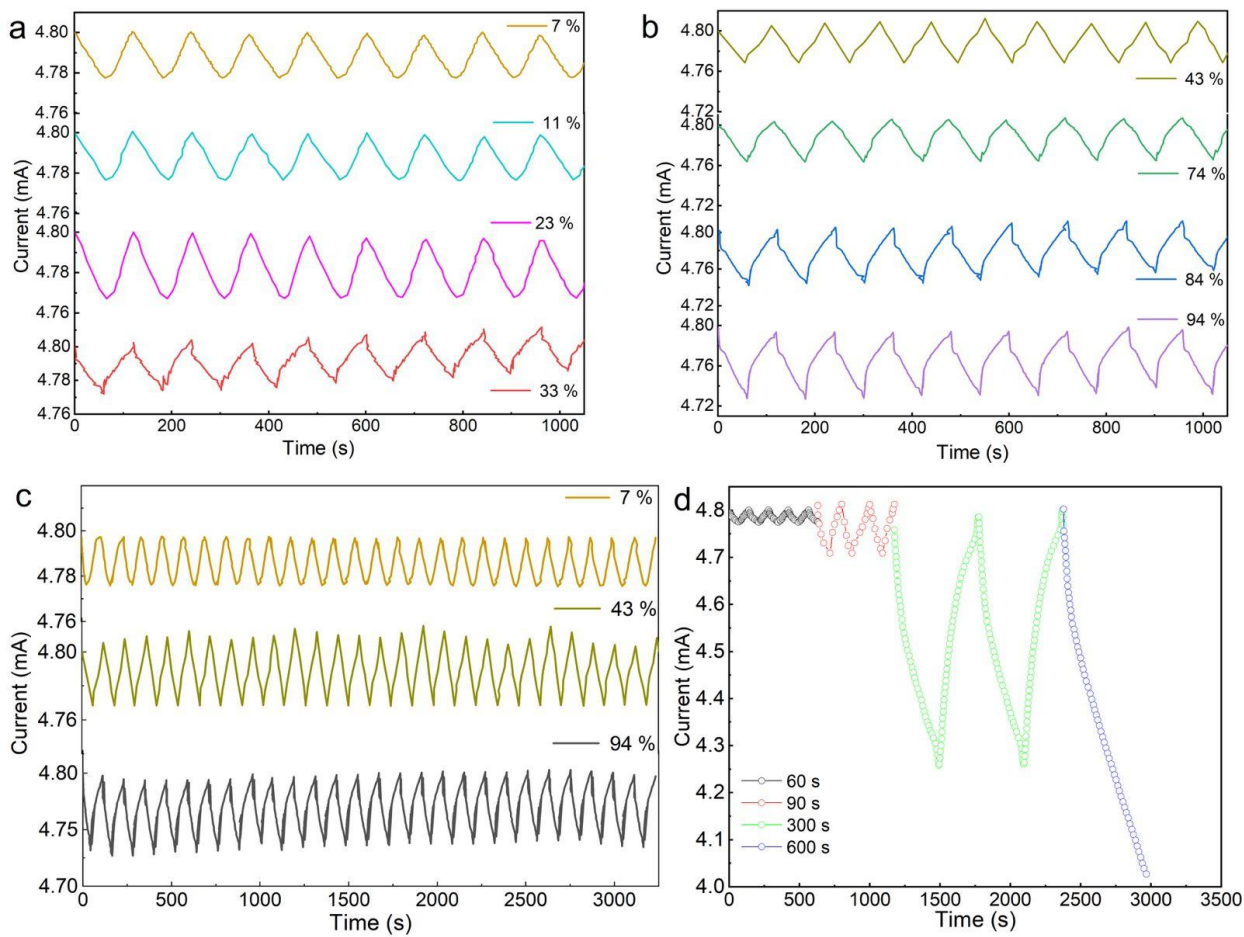

**Figure S6**. Raw humidity response data measured at RH levels of 7%, 11%, 23%, 33% (**a**), 43%, 75%, 84%, and 94% (**b**). Sensor's response over 3250 s for the lowest, intermediate and highest RH levels (**c**). Humidity response at 100% RH (using distilled water) for different exposure times (**d**). Measurements were performed at an applied voltage of 0.01 V and a temperature of 23°C.

**Table S3** Summary of humidity-sensing parameters at various RH levels, including sensitivity, response/recovery time, and signal amplitude.

| RH(%) | Sensitivity (%) | $t_{res}$ (s) | $t_{rec}$ (s) | ΔI (μA) |
|---|---|---|---|---|
| 7 | 0.46 | 45.83 | 34.26 | 21 |
| 11 | 0.48 | 46.87 | 34.35 | 21 |
| 23 | 0.49 | 46.24 | 34.11 | 27 |
| 33 | 0.68 | 46.89 | 44.72 | 27.5 |

| 43 | 0.65   | 46.6  | 43.34 | 40 |
| 75 | 0.74   | 51.01 | 43.45 | 41 |
| 84 | 1.16%  | 51.29 | 50.16 | 59 |
| 94 | 1.51%  | 52.47 | 47.29 | 64 |

Table S4. Sensing performance of the humidity sensor calculated by different equations.

| Equation | Sensitivity |
| --- | --- |
| $S(33\%) = \frac{I_{7\%} - I_{33\%}}{I_{33\%}} \times 100\%$  [S1] | 0.113 % |
| $S(33\%) = \frac{I_{33\%}}{I_{7\%}} \times 100\%$  [S2] | 99.8 % |
| $S(33\%) = \frac{I_{33\%}}{I_{baseline}} \times 100\%$  [S3] | 99.4 % |
| $S(33\%) = \frac{I_h - I_0}{I_0} \times 100\%$ | 0.68 % |

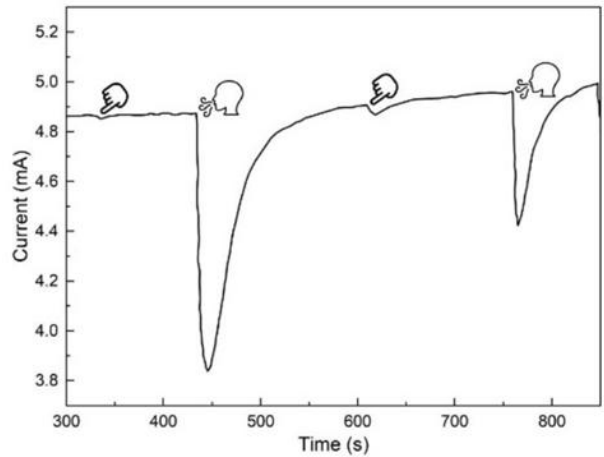

Figure S7. Skin humidity and exhaled breath of an adult measured by the sensor.

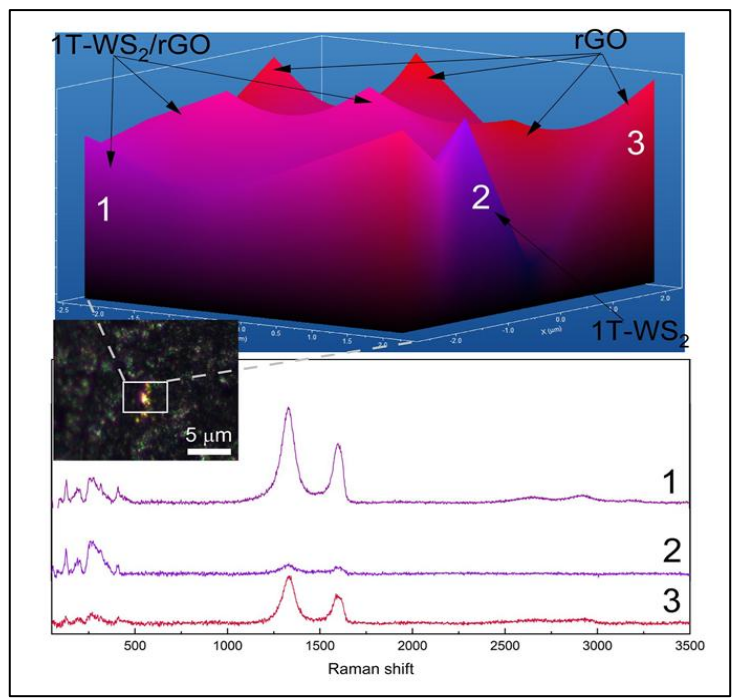

**Figure S8**. Raman mapping of the 1T-$WS_2$/rGO hybrid along with representative spectra (1-3). The optical image shows the area selected for mapping, approximately $2 \times 2\ \mu m^2$, acquired in point-by-point mode (hole size: 100). A 633 nm laser with a power of 0.5 mW was used, with an acquisition time of 5 s and 7 accumulations per point, to prevent the transition from the 1T to the 2H phase. The resulting Raman map was visualized based on the positions of characteristic peaks: blue indicates spectra dominated by 1T-$WS_2$ peaks, red corresponds to spectra dominated by rGO peaks, and purple/mixed colors highlight regions where both 1T-$WS_2$ and rGO peaks are simultaneously well-resolved and intense.